\makeatletter
\@ifundefined{@parse@version@dash}{%
\def\@parse@version#1{\@parse@version@0#1}
\def\@parse@version@#1/#2/#3#4#5\@nil{%
\@parse@version@dash#1-#2-#3#4\@nil}
\def\@parse@version@dash#1-#2-#3#4#5\@nil{%
  \if\relax#2\relax\else#1\fi#2#3#4 }
}{}
\makeatother
\documentclass[amsmath,twocolumn,amssymb,showkeys,superscriptaddress,aps,prd,10pt,twocolumn]{revtex4-2}
\usepackage{graphicx}
\usepackage[caption=false]{subfig}
\usepackage{multirow}
\usepackage{appendix}
\usepackage{graphicx,epstopdf}
\usepackage{subfig}

\usepackage{colortbl}
\usepackage{xcolor}
\usepackage[colorlinks=true, urlcolor=blue, linkcolor=blue, citecolor=blue]{hyperref}
\usepackage{float}
\usepackage{slashed}

\begin{document}

\title{Investigation of $\Delta^0\Delta^0$ dibaryon in QCD}

\author{H. Mutuk}
 \email{hmutuk@omu.edu.tr}
\affiliation{Department of Physics, Faculty of Arts and Sciences, Ondokuz Mayis University, Atakum, 55200 Samsun, T\"{u}rkiye}
\affiliation{Institute for Advanced Simulation, Institut f\"{u}r Kernphysik and J\"{u}lich Center for Hadron
Physics, Forschungszentrum J\"{u}lich, D-52425 J\"{u}lich, Germany}
\author{K.  Azizi}
\email{kazem.azizi@ut.ac.ir}
\thanks{Corresponding Author}
\affiliation{Department of Physics, University of Tehran, North Karegar Avenue, Tehran
14395-547, Iran}
\affiliation{Department of Physics, Do\v{g}u\c{s} University, Dudullu-\"{U}mraniye, 34775 Istanbul, T\"{u}rkiye}

\begin{abstract}
 We study a hypothetical $\Delta^0\Delta^0$ dibaryon state by means of QCD sum rule method. We construct a scalar interpolating current to extract the mass and decay constant of this state taking into account contributions of various quark, gluon and mixed vacuum condensates. The predictions for the mass and decay constants are $m_D=2326^{+114}_{-126} ~\mathrm{MeV}$  and $f_D=2.94^{+0.30}_{-0.34}\times 10^{-4} ~\mathrm{GeV}^7$, respectively. We also made an estimation about binding energy and size of the $\Delta^0\Delta^0$ dibaryon. The obtained mass value is below  the $\Delta^0\Delta^0$ threshold and may be a good dibaryon candidate for being observed experimentally. 
\end{abstract}

\maketitle
\section{\label{sec:level1}Introduction}
The possibility of multiquark states had been argued in the early days of the quark model \cite{Gell-Mann:1964ewy} in which multiquark states such as tetraquarks ($qq\bar{q}\bar{q}$) and pentaquarks ($qqqq\bar{q}$) were proposed apart from the conventional meson ($q\bar{q}$) and baryon ($qqq$, $\bar{q}\bar{q}\bar{q}$) states. In the past  two decades, experimental observation of the so-called ``exotic" hidden-charm $XYZ$ tetraquark and $P_c$ pentaquark states \cite{Chen:2016qju,Richard:2016eis,Lebed:2016hpi,Esposito:2016noz,Guo:2017jvc,Ali:2017jda,Olsen:2017bmm,Liu:2019zoy,Brambilla:2019esw,Agaev:2020zad} opened a new window in the quark model.

Another type of multiquark system is dibaryon or hexaquark. A dibaryon  consists of two  baryons. Deuteron is a well-known example of dibaryon. A recent dibaryon candidate  $d^\ast(2380)$ was observed at COSY \cite{WASA-at-COSY:2011bjg,WASA-at-COSY:2012seb,WASA-at-COSY:2013fzt,WASA-at-COSY:2014dmv}. This state was first predicted on the basis of SU(6) symmetry by Dyson and Xuong in 1964 \cite{Dyson:1964xwa}, although its existence was challenged by some studies (see for instance Refs. \cite{Ikeno:2021frl,Molina:2021bwp}). Another dibaryon structure was proposed by Jaffe \cite{Jaffe:1976yi}. This dibaryon was assumed to have $uuddss$ quark content and named dihyperon or mostly preferred the H-dibaryon. The mass and stability of H-dibaryon were investigated in several works followed by  Jaffe's paper such as the MIT bag Model \cite{Liu:1982wg}, the chiral model \cite{Yost:1985mj}, the quark cluster model \cite{Oka:1983ku}, the solitons in chiral quark model \cite{Balachandran:1985fb}, the nonrelativistic quark model \cite{Oka:1986fr}, the nonrelativistic quark cluster model \cite{Straub:1988mz,Koike:1989ak}, the QCD sum rule \cite{Larin:1985yt,Kodama:1994np} and lattice QCD \cite{Inoue:2010es}. The results of these studies are controversial; mass results change from 1.1 GeV to 2.2 GeV. Furthermore, some of these studies pointed out that $q_1q_1q_2q_2q_3q_3$ type dibaryons are most favorable for being stable. These dibaryon states were studied by means of stability in \cite{Leandri:1997ge,Vijande:2011im}. Ref. \cite{Leandri:1997ge} used a simple chromomagnetic model to study dibaryon states. No stable dibaryon state is found but it is mentioned that some dibaryons can be seen as resonances in certain channels. On the other hand, a QCD inspired string model calculation showed that ground state of   a six-quark ($q^6$) state is stable against dissociation into two baryons \cite{Vijande:2011im}. In a recent study, spectroscopic parameters of $S=uuddss$ hexaquark were calculated and its candidacy for dark matter was argued \cite{Azizi:2019xla}. Ref. \cite{Beiming:2021bkj} used the extension of the  G\"{u}rsey-Radicati mass formula to obtain some nonstrange dibaryon mass spectrum.

The strangest dibaryon $\Omega \Omega$ was studied by lattice QCD and was found to be located near the unitary regime \cite{Gongyo:2017fjb}. This lattice QCD simulation triggered further studies. In Ref. \cite{Morita:2019rph}, momentum correlation functions of $\Omega \Omega$ and $p\Omega$ dibaryons are investigated via constructing correlation function by single-particle distributions. Quark delocalization color screening and chiral quark models are used to reanalyze $\Omega \Omega$ dibaryon states with $J^P=0^+,1^-,2^+,3^-$ quantum numbers and $J^P=0^+$ is found to be bound \cite{Huang:2019hmq}.  $\Omega \Omega$ dibaryon states with $J^P=0^+$ and $J^P=2^+$ are investigated in a molecular picture by using QCD sum rule \cite{Chen:2019vdh}. An evidence is found for a strangeness $S=-1$ $\Sigma N$ dibaryon bound state in the $\Lambda N$ cross section  \cite{Haidenbauer:2021smk}. The $N \Omega$ system was studied in the $^5S_2$ channel by lattice QCD and a quasi-bound state was predicted near unitarity \cite{HALQCD:2018qyu}. The possible $N\Omega_{ccc}$ and $N\Omega_{bbb}$ dibaryons  are studied in quark delocalization color screening model \cite{Huang:2019esu}. Their results showed that these states are bound. $N\Omega$ systems in $^3S_1$ and $^5S_2$ channels with quantum numbers of $J^P=1^+$ and $J^P=2^+$ are studied in the molecular picture by QCD sum rule  \cite{Chen:2021hxs}. They found that $N\Omega$ dibaryon may exist in the  $^5S_2$ channel. In Ref. \cite{HALQCD:2019wsz}, the S-wave  $\Lambda \Lambda$ and  $N\Xi$ systems are studied by lattice QCD. It was found that the $\Lambda \Lambda$ interaction at low energies resulted in a weak interaction whereas $N\Xi$ interaction leaded the $N\Xi$ system near unitarity. 

Dibaryons with two or more heavy quark contents are also being studied. In Ref. \cite{Wang:2017sto}, the doubly charmed scalar $uuddcc$ hexaquark state is studied by QCD sum rule method and $6.60^{+0.12}_{-0.09}~\mathrm{GeV}$ mass value is obtained. The spectra of the hidden-bottom and -charm hexaquark states were investigated in molecular picture via QCD sum rule \cite{Wan:2019ake}. Their results indicate that $0^{++}$ and $1^{--}$ states in the $b$ quark sector are possible baryonium states. The triply-charmed $\Xi_{cc}\Sigma_c$ dibaryon states were also studied via QCD sum rule by considering dibaryon and two-baryon scattering states \cite{Wang:2019gal}. $\Sigma_c \Xi_{cc}$, $\Omega_c \Omega_{cc}$, $\Sigma_b \Xi_{bb}$, $\Omega_b \Omega_{bb}$, and $\Omega_{ccb} \Omega_{cbb}$ particles were studied by lattice QCD formalism \cite{Junnarkar:2019equ}. Among these states, evidence for the existence of $\Sigma_c \Xi_{cc}$ and $\Sigma_b \Xi_{bb}$ dibaryons could not be obtained whereas for the other dibaryons their presence is suggested. Triply-charmed $\Xi_{cc}\Sigma_c$ dibaryon states were investigated in the one-boson exchange approach \cite{Pan:2020xek}. A number of triply charmed and triply bottomed states of isospins $1/2$ and $3/2$ are found to be bound. In  Ref. \cite{Wang:2020jqu}, the axial vector triply-charmed hexaquark state is studied with the QCD sum rule method. The possibility of $bbbccc$ dibaryon was investigated by a constituent quark model \cite{Richard:2020zxb}. They observed that there is no evidence for any stable type of $bbbccc$ dibaryon. The existence of fully heavy dibaryons $\Omega_{ccc}\Omega_{bbb}$, $\Omega_{ccc}\Omega_{ccc}$ and $\Omega_{bbb}\Omega_{bbb}$ with $J = 0, 1, 2, 3$ and $P = ±1$ quantum numbers is investigated in constituent quark model \cite{Huang:2020bmb}. Their results showed that  all-heavy $ b $ or $ c $ dibaryons of $J^P = 0^+$ are possible to be bound whereas $cccbbb$ configuration is not. Scattering properties of $\Omega_{ccc}\Omega_{ccc}$ dibaryon system were studied by lattice QCD method and found to be located in the unitarity regime \cite{Lyu:2021qsh}. Low-lying doubly-heavy dibaryon systems with strangeness $S=0$ and various quantum numbers were studied in the quark delocalization color screening framework \cite{Xia:2021nif}. They found three bound state candidates for doubly-charm and doubly-bottom dibaryon systems. In Ref. \cite{Liu:2021pdu}, prediction of  $\Omega_{bbb} \Omega_{bbb}$ dibaryon state was studied in the extended one-boson exchange model and yielded a possible $\Omega_{ccc} \Omega_{ccc}$ dibaryon state. 

In addition to dibaryon configurations mentioned above, according to quantum chromodynamics (QCD), many other dibaryon structures are possible. The theory of QCD and its Lagrangian, was first proposed by Harald Fritzsch and Murray Gell-Mann in 1972 \cite{Fritzsch:1972jv}. In principle, besides the dynamics of quarks and gluons, this Lagrangian should be responsible for hadrons and determination of their properties. Unfortunately, it is valid only in a limited region. Today, QCD is widely accepted as a true theory of the strong interactions, despite the fact that it is not directly applicable at the low-energy region. Hadron-hadron interactions are prototypes in understanding QCD as a fundamental theory of strong interactions. However, for hadron-hadron interactions and exotic systems, using QCD directly to probe low-energy region is difficult due to the nonperturbative aspects. Therefore some nonperturbative methods are required. QCD sum rule, lattice QCD, chiral perturbation theory, quark model and effective field theories are in play to study the multiquark systems like tetraquarks and pentaquarks from the physical perspective. Among these nonperturbative methods, the  QCD sum rule is a powerful nonperturbative tool to study many hadronic phenomena.  The QCD sum rule method is based on the QCD Lagrangian which is the main advantage of the method. It is relativistic and does not include any free parameter. 

In this present study, we compute mass and decay constant of $\Delta^0\Delta^0$ dibaryon, which is a possible dibaryon candidate in the light quark sector by using the QCD sum rule method. This possible dibaryon state was studied via extended chiral SU(3) quark model \cite{Dai:2005kt}, the quark delocalization color screening and  the chiral quark models \cite{Huang:2013nba}, and three-body interaction model \cite{Gal:2013dca}. To the best of our knowledge, there is no reported QCD sum rule study for this dibaryon.

This work is arranged as follows: in Section \ref{sec:level2} we briefly introduce the  QCD sum rule method and obtain sum rules for spectroscopic parameters of the $\Delta^0\Delta^0$ dibaryon. The calculations were done by carrying out various vacuum condensates. In Section \ref{sec:level3}, we present numerical results and discussions. Section \ref{sec:level4} is reserved for our concluding remarks.

\section{\label{sec:level2}Methodology}
The QCD sum rule (QCDSR) is a semiphenomenological framework to extract spectroscopic information from the QCD Lagrangian. The method was first proposed for mesons in \cite{Shifman:1978bx,Shifman:1978by} and later generalized to baryons \cite{Ioffe:1981kw}.
One can relate the hadron spectrum to the QCD Lagrangian in this method. QCDSR handles bound state problem by starting to study short distance relations and then moving to large distance relations by including nonperturbative effects of QCD vacuum. This is done via an appropriate correlation function. 

QCDSR formalism handles the corresponding state by starting from short distances (large $q^2$) where asymptotic freedom makes perturbative calculation doable and moving to large distances (low $q^2$) where hadronic states are formed due to the nonperturbative interactions. Provided that there is some interval in $q$, which is the fundamental hypothesis of the QCDSR formalism, where both  representations overlap, one can obtain information about hadronic properties. Masses and decay constants (residues) of the hadrons are generally investigated within the two-point correlation functions, whereas the strong coupling constants and form factors of different interactions  are studied within three-point or light-cone correlation functions. 
 
In order to obtain sum rules for mass and decay constant, we study the two-point QCDSR formalism. The method focuses on the correlation function
\begin{equation}
\Pi (p)=i\int d^{4}xe^{ipx}\langle 0|\mathcal{T}\{J(x)J^{\dagger
}(0)\}|0\rangle,  \label{eq:CorrF}
\end{equation}%
and obtains sum rules to compute physical properties of the related state. Here, $\mathcal{T}$ is the time order operator and $J(x)$ is the corresponding interpolating current. To study the $\Delta^0 \Delta^0$ dibaryon in the framework of QCDSR, we need to construct the interpolating current which is the main ingredient of the analysis. We construct $\Delta^0 \Delta^0$ interpolating current by using Ioffe current for the $\Delta^0$ baryon as
\begin{eqnarray}\label{current}
\eta_{\mu}=\frac{1}{\sqrt{3}}\varepsilon^{abc}\left[\vphantom{\int_0^{x_2}}2(d^{aT}C\gamma_{\mu}u^{b})d^{cT}+(d^{aT}C\gamma_{\mu}d^{b})u^{c}\right],
\end{eqnarray}
where $C$ is the charge conjugation operator and $a,b,c$ are color indices. Then, the $J^P=0^+$ interpolating current for $\Delta^0 \Delta^0$ dibaryon can be constructed as 
\begin{equation}
J(x)=\eta_{\mu} \cdot C \gamma_5  \cdot \eta^{\mu}.
\end{equation}

There are two paths to calculate correlation function in the standard prescriptions of the QCDSR method. At the first path which is called phenomenological side, the correlation function is being calculated by inserting intermediate hadronic states that have the same quantum numbers of interpolating current $J(x)$. At this step, correlation function is represented by the hadronic degrees of freedom. To do this, we express the correlation function (Eq. \ref{eq:CorrF}) with the help of $\Delta^0 \Delta^0$ dibaryon mass $m_{D}$, coupling $f_D$, and its matrix element
\begin{equation}
\langle 0 \vert J \vert D  \rangle=m_D f_D. \label{matele}
\end{equation}
The ground state contribution can be isolated from the contributions of higher states and continuum for $\Pi (p)$. Then we obtain

\begin{equation}
\Pi^{\text{Phen}}(p)=\frac{\langle 0 \vert J \vert D(p)\rangle \langle D(p) \vert J^\dagger \vert 0 \rangle}{m_D^2-p^2} + \cdots .
\end{equation}
With the definition of matrix element (Eq. \ref{matele}), we can rewrite  $\Pi^{\text{Phen}}(p)$ as
\begin{equation}
\Pi^{\text{Phen}}(p)=\frac{m_D^2 f_D^2}{m_D^2-p^2}+ \cdots, \label{CorP}
\end{equation}
where $\cdots$ represents contributions of higher states and continuum.

In the second step, calculation of the correlation function (Eq. \ref{eq:CorrF}) proceeds by operator product expansion (OPE). This representation is called OPE or QCD side of the correlation function. We insert the interpolating current $J(x)$ which is composed of the quark fields into the correlation function and contract all the quark fields with the help of Wick theorem. As a result we obtain 
\begin{eqnarray}
\Pi^{\text{QCD}}(p) &=& \frac{16}{9} \varepsilon^{abc} \varepsilon^{def}\varepsilon^{a^\prime b^\prime c^\prime} \varepsilon^{d^\prime e^\prime f^\prime}\int d^4x e^{ipx}   \nonumber \\ &\times& \{ \mathrm{Tr}\left[ \gamma_{\mu} S_u^{bb^\prime}(x)\gamma_{\nu} \tilde{S}_d^{a^\prime a}(-x) \right]  \nonumber \\
&\times & \mathrm{Tr}\left[ \gamma_{\mu} S_u^{ee^\prime}(x)\gamma_{\nu} \tilde{S}_d^{d^\prime d}(-x) \right] \nonumber \\ 
&\times & \mathrm{Tr}\left[ \gamma_{5} S_d^{ff^\prime}(x)\gamma_{5} \tilde{S}_d^{cc^\prime}(-x) \right] \}  \nonumber \\
&+&\mbox{many similar terms}, \label{corOp}
\end{eqnarray}
where $\tilde{S}(x)=CS^T(x)C$. To proceed in  the QCD side, the light $u$ and $d$ quarks propagators are used:
\begin{eqnarray}
&&S_{q}^{ab}(x)=i\frac{\slashed x}{2\pi ^{2}x^{4}}\delta _{ab}-\frac{m_{q}}{%
4\pi ^{2}x^{2}}\delta _{ab}-\frac{\langle \overline{q}q\rangle }{12}\left(
1-i\frac{m_{q}}{4}\slashed x\right) \delta _{ab}  \notag \\
&&-\frac{x^{2}}{192}\langle \overline{q}g_{s}\sigma Gq\rangle \left( 1-i%
\frac{m_{q}}{6}\slashed x\right) \delta _{ab}  \notag \\
&&-\frac{ig_{s}G_{ab}^{\mu \nu }}{32\pi ^{2}x^{2}}\left[ \slashed x\sigma
_{\mu \nu }+\sigma _{\mu \nu }\slashed x\right] -\frac{\slashed %
xx^{2}g_{s}^{2}}{7776}\langle \overline{q}q\rangle ^{2}\delta _{ab}  \notag
\\
&&-\frac{x^{4}\langle \overline{q}q\rangle \langle g_{s}^{2}G^{2}\rangle }{%
27648}\delta _{ab}+\frac{m_{q}g_{s}}{32\pi ^{2}}G_{ab}^{\mu \nu }\sigma
_{\mu \nu }\left[ \ln \left( \frac{-x^{2}\Lambda ^{2}}{4}\right) +2\gamma
_{E}\right]  \notag \\
&&+\cdots,  \label{eq:LQProp}
\end{eqnarray}%
where $q=u,~d$, $\gamma _{E}\simeq 0.577$ is the Euler constant, $\Lambda $ is a scale parameter, $G_{ab}^{\mu \nu}\equiv G_{A}^{\mu \nu }t_{ab}^{A},\ A=1,2,\cdots, 8$, and $t^{A}=\lambda^{A}/2$, with $\lambda ^{A}$ being the Gell-Mann matrices.

We shall note that we use the factorization approximation for the higher dimension operators.  The central idea behind factorization hypothesis is replacing vacuum expectation values of higher dimensional operators by the products of the lower dimensional ones. This is one of the key sources of uncertainties in the QCDSR approach.  For example the four-quark condensate ($d=6$) $\langle \bar{q} q \bar{q} q \rangle$ and quark condensate times mixed condensate ($d=8$) $\langle \bar{q} q \bar{q} g \sigma \cdot G q \rangle$ can be factorized as
\begin{eqnarray}
\langle \bar{q} q \bar{q} q \rangle &=& \rho \langle \bar{q} q \rangle^2, \nonumber \\
 \langle \bar{q} q \bar{q} g \sigma \cdot Gq \rangle &=& \rho \langle \bar{q} q \rangle \langle \bar{q} g \sigma \cdot Gq \rangle 
\end{eqnarray}
where $\rho$ is a parameter and account deviations from the hypothesis.  $\rho=1$ gives the vacuum saturation  and $\rho=2.1$ indicates the violation of the factorization hypothesis \cite{Narison:1989aq,Launer:1983ib,Narison:2009vy,Braghin:2014nva}. It was shown in  Ref. \cite{Albuquerque:2010fm} that $d=8$ condensate under factorization hypothesis is almost negligible by using a molecular current couple to $D_s^\ast D_{s0}^\ast$. The effect of nonfactorized diagrams is found to be small for doubly hidden $0^{++}$ molecule and tetraquark states in Ref. \cite{Albuquerque:2021erv}. The effect of factorization in a sum rule calculation at LO is about $ 2.2\% $  for the residue (decay constant), and $0.5\%  $  for the mass in the $XYZ$ exotic states \cite{Albuquerque:2016znh,Albuquerque:2017vfq}.

In order to extract information from correlation function, the invariant amplitudes $\Pi^{\text{Phen}}(p)$ and $\Pi^{\text{QCD}}(p)$ must be equated. To suppress the contributions of the higher states and continuum, the Borel transformation should be applied to both sides of the obtained sum rule. In this step, the  quark-hadron duality assumption is used to perform continuum subtraction. One can reach an expression, after these technical elaborations, in terms of  the mass $m_D$ and decay constant $f_D$ of the $\Delta^0\Delta^0 $ in the hadronic side and the QCD degrees of freedom like quark masses, vacuum condensates of different dimensions, etc in the QCD side. 
After Borel transformation and continuum subtraction the QCD side of the sum rule takes the  following form:

\begin{equation}
\tilde{\Pi}^{\text{QCD}}=\int_{\mathcal{M}^2}^{s_0} \rho(s) e^{-s/M^2} ds  + \Gamma, \label{rhogam}
\end{equation}
where $\mathcal{M}=(4m_d+2m_u)$, $s_0$ is the continuum threshold which is the energy  characterizing beginning of the continuum, $M^2$ is the Borel parameter and $\rho(s)$ is the spectral density obtained from the imaginary part of the correlation function, $\rho(s)=\text{Im}[\Pi(s)]/\pi$. The explicit expressions of the $\rho(s)$ and $\Gamma$ under the assumptions of $m_u \to 0$ and $m_d \to 0$ are given as 
\begin{widetext}
\begin{eqnarray}
\rho(s)&=& \frac{1873 s^7}{2^{22}3^5 5^2 7^2\pi^{10}}-m_0^2 \langle \bar{q}q \rangle^4 \left( \frac{3715457 g_s^2}{2^{12} 3^8 \pi^4} + \frac{61091}{2^{10}3^2 \pi^2}\right)+ \langle \bar{q}q \rangle^4 s \left(\frac{9365g_s^4 }{2^8 3^{10}\pi^6}+  \frac{4771 g_s^2}{2^3 3^8\pi^4} + \frac{4771}{2^3 3^8\pi^2}  \right)\nonumber \\ &+& \langle \bar{q}q \rangle^2 \left( \frac{243193 m_0^4 s^2}{2^{18}3^3 \pi^6}- \frac{1034923 m_0^2 s^3}{2^{18}3^5 \pi^6} + \frac{1873 g_s^2 s^4}{2^{14}3^7 5 \pi^8}+ \frac{4771 s^4}{2^{12}3^5 5 \pi^6} \right)  \nonumber \\ 
&+& \langle g^2 G^2 \rangle \left(\frac{1159m_0^4 \langle \bar{q}q \rangle^2}{2^{20}3^5 \pi^6} - \frac{2347m_0^2 \langle \bar{q}q \rangle^2 s}{2^{19}3^5 \pi^6}+ \frac{g_s^2\langle \bar{q}q \rangle^2 s^2}{2^{12}3^5 \pi^8} + \frac{11 \langle \bar{q}q \rangle^2 s^2}{2^{16}3^3 \pi^6} + \frac{s^5}{2^{21}3^2 5^2 \pi^10} \right),
\end{eqnarray}
\begin{eqnarray}
\Gamma &=& m_0^4 \langle \bar{q}q \rangle^4 \left(\frac{2494081 g_s^2}{2^{14} 3^8 \pi^4} + \frac{1038769}{2^{12} 3^4 \pi^2} \right) + \langle g^2 G^2 \rangle \langle \bar{q}q \rangle^4 \left(\frac{g_s^4 }{2^9 3^8 \pi^6} + \frac{11 g_s^2}{2^{11} 3^7 \pi^4} + \frac{59}{2^{10}3^6 \pi^2} \right).
\end{eqnarray}
\end{widetext}

Thus, the following sum rule relates the hadronic parameters to the QCD degrees of freedom as well as the auxiliary parameters $M^2$  and $ s_0 $:
\begin{equation}
f_D^2 m_D^2 e^{-\frac{m_D^2}{M^2}}=\tilde{\Pi}^{\text{QCD}}.
\end{equation}
The mass $m_D$ and  residue $f_D$  of the dibaryon are obtained from the above sum rule as
\begin{equation}
m_D^2 (s_0, M^2)=\frac{\tilde{\Pi}^{\prime \text{QCD}}}{\tilde{\Pi}^{\text{QCD}}},
\end{equation}
with $\tilde{\Pi}^{\prime \text{QCD}}=\frac{d}{d(-\frac{1}{M^2})} \tilde{\Pi}^{\text{QCD}}$,
and 
\begin{equation}
f_D^2 (s_0, M^2)  = \frac{e^{m_D^2/M^2}}{m_D^2} \tilde{\Pi}^{\text{QCD}}.
\end{equation}

\section{\label{sec:level3}Numerical Results and Discussion}
We numerically analyze the sum rules for the $\Delta^0 \Delta^0 $ by using the input parameters which are given in Table \ref{tab:table1}.

\begin{table}[H]
\caption{\label{tab:table1}QCD input parameters}
\begin{ruledtabular}
\begin{tabular}{lc}
Parameter&Numerical value\\
\hline
$\langle \bar{q} q \rangle$ & $(-0.24 \pm 0.01)^3 ~ \text{GeV}^3$ \cite{Belyaev:1982sa}\\
$m_0^2$ & $(0.8 \pm 0.1) ~ \text{GeV}^2$ \cite{Belyaev:1982sa} \\
$\langle \frac{\alpha_s}{\pi} G^2 \rangle$ & $(0.012 \pm 0.004) ~ \text{GeV}^4$  \cite{Belyaev:1982cd}\\
\end{tabular}
\end{ruledtabular}
\end{table}

There are two further parameters which are used in QCDSR analysis: the $s_0$ continuum threshold parameter which arises after continuum subtraction via using quark-hadron duality and the $M^2$ Borel parameter which arises after Borel transformation as we also noted before. Any physical quantity should be independent of these parameters. Although these parameters are auxiliary, they are not arbitrary and should verify standard constraints. In order to extract reliable results from QCDSR, we must find the working regions of $s_0$ and $M^2$ that the results have $s_0$ and $M^2$ stability. So an important problem emerges for the proper choice of Borel $M^2$ and continuum threshold $s_0$ parameters. 

For Borel parameter $M^2$, a good OPE convergence and suppression of higher and continuum states are necessary to determine the region. The upper and lower limits of Borel parameter need an additional examination of the sum rules. 

At the upper limit of Borel parameter, the pole contribution (PC) should reside fairly in the correlation function whereas at the lower limit of the Borel parameter, it must be a dominant contribution. This examination can be done by defining PC as
\begin{equation}
\text{PC}=\frac{\tilde{\Pi}^{\text{QCD}}(s_0,M^2)}{\tilde{\Pi}^{\text{QCD}}(s_0=\infty,M^2)}.
\end{equation}

A balance can be possible between upper limit $M^2_{\text{max}}$ and lower limit $M^2_{\text{min}}$ of the Borel parameter, which provides both convergence of the series and dominance of the single resonance contribution.

The continuum threshold $s_0$ isolates the ground state contribution from the continuum states and higher resonances in the correlation function. $s_0$ characterizes the threshold where continuum states begin. In a standard analysis, $s_0$ should be just below  the energy of the first excited state. The energy that is needed for this excitation reads as $\delta=\sqrt{s_0}-m$, with $m$ being the mass of ground state. In order to estimate $s_0$ value, experimental information on the masses of the ground and first excited states of the channel under study is  required. For the conventional hadrons, parameters of excited states can be obtained from either experimental measurements or theoretical studies. For these states $ \delta $ varies mainly  in the interval, $0.3~ \text{GeV} \leq \delta\leq 0.8~ \text{GeV} $.  For multiquark hadrons or better to say exotic states, however,  there can be lack of relevant information. In this case $s_0$ can be obtained by limits  imposed on PC and convergence of OPE. The interval for the $s_0$ can be chosen as the smallest value which provides a reliable Borel region. One can fix $s_0$, by satisfying other constraints,  to achieve a maximum for PC. 
Equipped with this it is possible to monitor numerical consistency of analytical calculations. 
Performed analysis shows that the working windows are as 
\begin{equation}
M^{2}\in [2.4,3.0]~\mathrm{GeV}^{2},\ s_{0}\in [8,10]~\mathrm{GeV}^{2},  \label{workreg}
\end{equation}%
that meet all aforementioned restrictions. As it is clear, the working window for $ s_0 $ lies much higher than the production threshold of two $ \Delta^0 $
baryons ($  4m_{\Delta^0}^2\simeq 6.07~\mathrm{GeV}^{2}$). The obtained interval for continuum threshold corresponds to $(m_{\Delta^0\Delta^0}+\delta)^2$ with $0.50~ \text{GeV} \leq \delta\leq 0.84~ \text{GeV} $, which is a reasonable interval of energy to excite a $\Delta^0\Delta^0$  system to its first excited state.  Note that in $ \Delta^0 $ three-quark baryon channel the energy difference between the ground and first excited state with the same quantum numbers, $ \Delta(1600)$,   is about $ 0.37~ \text{GeV} $.

Using the above  working regions for the auxiliary parameters, the pole contribution varies in the interval
\begin{equation}
0.12 \leq \mathrm{PC}\leq 0.68.
\end{equation}
In the standard analysis of QCDSR, the pole contribution should be larger than 1/2 for conventional hadrons (baryons and mesons). In the case of four-quark states, it turns out to be as $\mathrm{PC\geqslant 0.2}$. In Refs. \cite{Chen:2014vha,Chen:2019vdh}, it is pointed out that dibaryon spectral densities led to small PC and therefore failed to specify the Borel parameter.

To extract the mass $m_D$ and decay constant $f_D$ of the $\Delta^0\Delta^0$ dibaryon, we calculate them at different choices of the Borel parameter $M^2$ and continuum threshold $s_0$ and find their mean values averaged over the working regions given in Eq. (\ref{workreg}). Our results for $m_D$ and  $f_D$ read
\begin{eqnarray}
m_D &=& 2326^{+114}_{-126} ~\mathrm{MeV}, \nonumber \\
f_D &=& 2.94^{+0.30}_{-0.34} \times 10^{-4} ~\mathrm{GeV}^7. \label{results}
\end{eqnarray}
The values in Eq. (\ref{results}) correspond to average value of lower, middle, and upper values in the working regions. The plus and minus bounds for mass and decay constant yielded from highest and lowest values of $m_D$ and $f_D$. In Figs. \ref{fig:Mass} and \ref{fig:decayconstant}, we plot the mass and decay constant of the $\Delta^0\Delta^0$ dibaryon as functions of $M^2$ and $s_0$.

\begin{widetext}

\begin{figure}[h!]
\begin{center}
\includegraphics[totalheight=6cm,width=8cm]{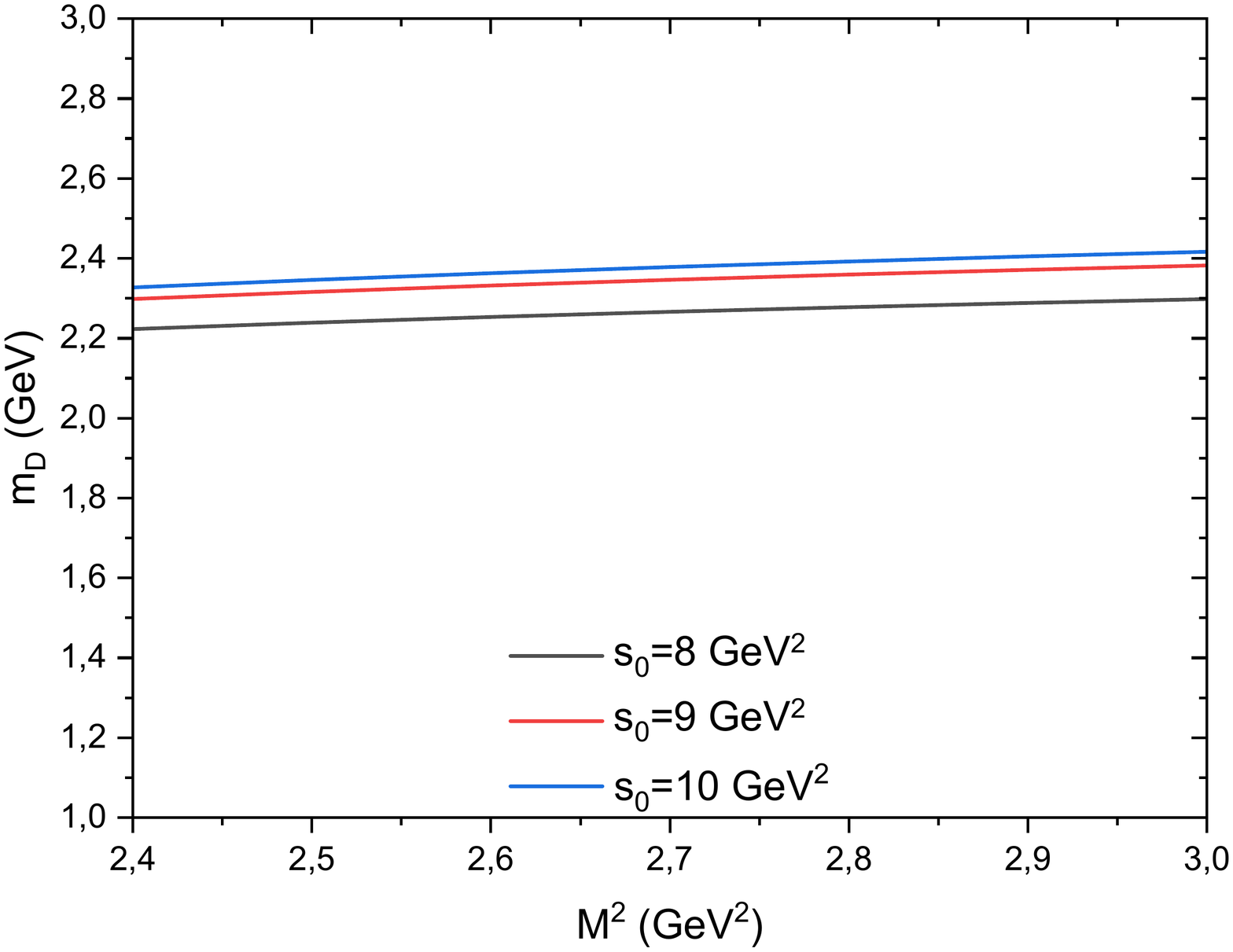}
\includegraphics[totalheight=6cm,width=8cm]{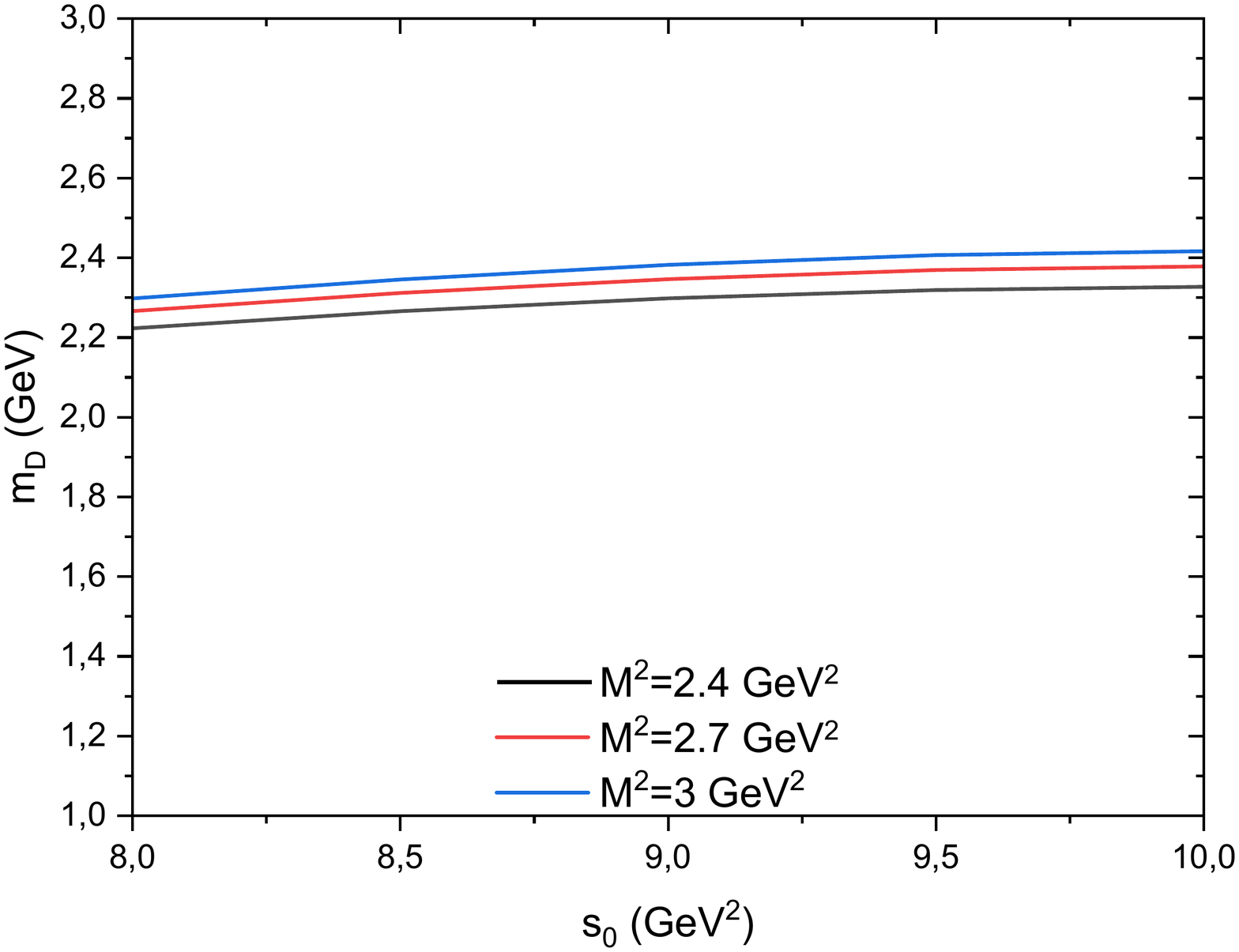}
\end{center}
\caption{The mass of the $\Delta^0\Delta^0$ dibaryon as a function of $M^{2}$ at fixed $s_{0} $ (left panel), and as a function of $s_0$ at fixed $M^2$ (right panel).}
\label{fig:Mass}
\end{figure}
\begin{figure}[h!]
\begin{center}
\includegraphics[totalheight=6cm,width=8cm]{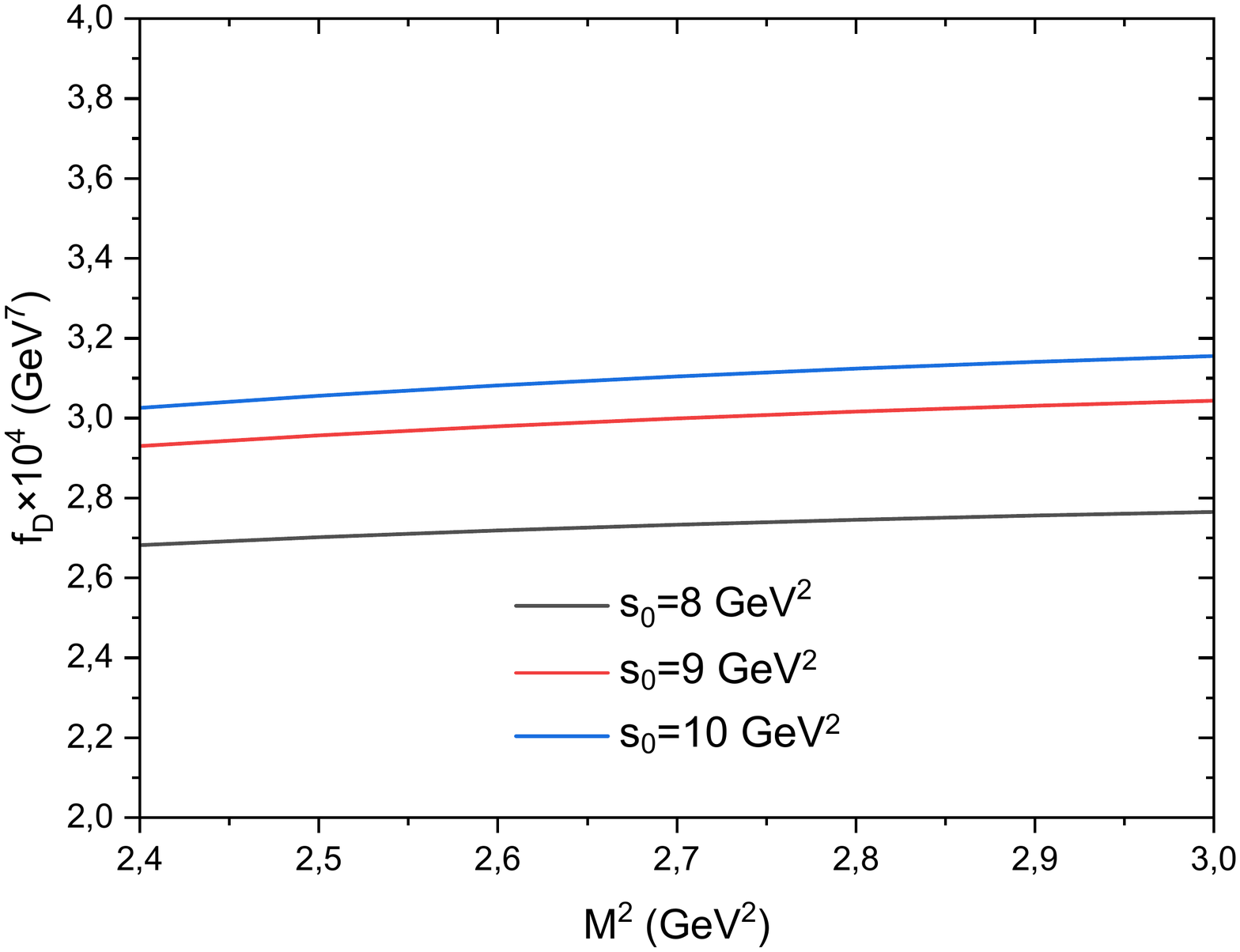}\,\, %
\includegraphics[totalheight=6cm,width=8cm]{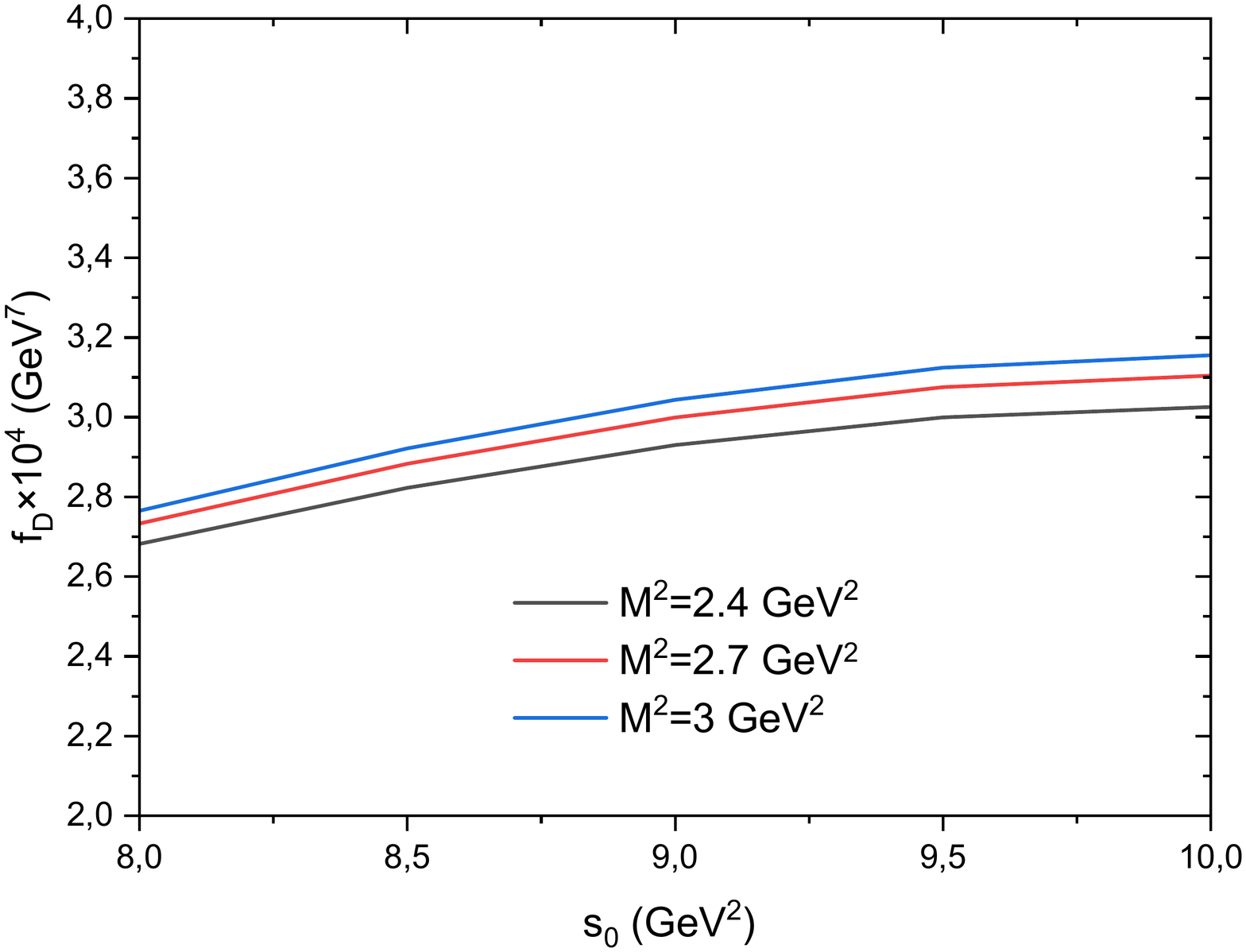}
\end{center}
\caption{Same as in Fig. \ref{fig:Mass}, but for the coupling $f_D$ of the $\Delta^0\Delta^0$ dibaryon.}
\label{fig:decayconstant}
\end{figure}

\end{widetext}

Fig. \ref{fig:Mass} (left panel) shows that the mass is relatively stable in the region for Borel parameter $M^2$. In right panel of Fig. \ref{fig:Mass}, the mass has a mild dependence on the region for continuum threshold $s_0$. In Fig. \ref{fig:decayconstant}, it can be seen from both left and right panels that there is a residual dependence on the parameters of $s_0$ and $M^2$. The dependence on $M^2$ and $s_0$ is essential part of theoretical errors in sum rule calculations. We have also added, on average,   $ 1 \% $ and  $ 3 \% $  uncertainties to the values of the mass and decay constant, respectively  due to the factorization hypothesis of the higher dimensional operators.    Theoretical uncertainty in the extracted mass $m_D$ is about $ \pm 5.5 \% $, where in the decay constant $f_D$, it is $ \pm 12 \% $. These uncertainties are well below than the accepted limits in QCDSR calculations. 
We shall remark that, in the present study, we do not consider the contributions of the radiative corrections. Although, the mass, due to the ratios of two sum rules,  is known to be less sensitive to these corrections,  the numerical value
of the residue obtained from one sum rule may be changed, considerably.   Note that, the NLO perturbative corrections to the sum rules of baryons were obtained to be  large \cite{Krasnikov:1982ea,Jamin:1987gq,Groote:1999zp}.  The NLO $\alpha_s$ corrections to the perturbative term in the OPE of spectral functions were found to be large for light tetraquark currents as well, however, these corrections to the whole sum rules are small because of the dominance of the nonperturbative condensate contributions \cite{Groote:2014pva}. In Ref. \cite{Groote:2006sy}, the NLO corrections to the spectral density of the light pentaquarks are found to be large and spoil momentum space QCD sum rule analysis, suggesting that a sum rule analysis is more reasonable in coordinate space. For pentaquark interpolating currents, the  NLO corrections to the correlators were also found to be large in Ref. \cite{Groote:2011my}. However, vacuum condensates at the order of $\mathcal{O}(\alpha_s^k)$ with $3/2 < k \leq 3$ play tiny roles in the sum rules for $\Lambda_c \Lambda_c$ dibaryon  according to Ref. \cite{Wang:2021qmn}. One should examine the contributions of the radiative corrections to the spectroscopic parameters of the light dibaryons.

The binding energy $B_E$ can be obtained via the relation
\begin{equation}
B_E=2m_{\Delta^0}-m_D,
\end{equation} 
where $m_{\Delta^0}$ is the mass of $\Delta^0$ baryon and $m_D$ is the predicted mass. The binding energy of $\Delta^0\Delta^0$ dibaryon is obtained to be $B_E \simeq 138~\mathrm{MeV}$ where $\Delta^0$ baryon mass is borrowed from Ref. \cite{Zyla:2020zbs}. Finally it will be interesting to look at the size of $\Delta^0\Delta^0$ dibaryon. The following relation \cite{Guo:2017jvc} can be used for this purpose
\begin{equation}
r \sim \frac{1}{\sqrt{2 \mu B_E}}
\end{equation}
where $r$ is the distance between the components and $\mu=m_1m_2/(m_1+m_2)$ denotes the reduced mass of the two-hadron system. We obtain the size as $r\simeq 0.48~\mathrm{fm}$. This is less than the so-called confinement radius, $1~\mathrm{fm}$.

\section{\label{sec:level4} Summary and Final Notes}
In the present work, we  considered the exotic scalar $\Delta^0\Delta^0$ dibaryon composed of $uddudd$ quarks. Using the QCDSR technique and by satisfying its requirements such as the pole dominance and OPE convergence, the values of its mass and decay constant (residue) were extracted as $m_D=2326^{+114}_{-126} ~\mathrm{MeV}$  and $f_D=2.94^{+0.30}_{-0.34}\times 10^{-4} ~\mathrm{GeV}^7$. 
 We obtained the  binding energy  $B_E \simeq 138~\mathrm{MeV}$ for the $\Delta^0\Delta^0$  system. We also made an estimation for the size of the $\Delta^0\Delta^0$ dibaryon which is $r \simeq 0.48~\mathrm{fm}$.
A QCDSR calculation can provide  evidence in favor of or against the existence of a state. The obtained result for the binding energy shows that the scalar  $\Delta^0\Delta^0$ dibaryon  system can be stable and  long-lived. We hope that our results will  be useful for both the future theoretical and the experimental studies on dibaryons. 

\begin{acknowledgments}
The work of H. Mutuk is partially supported by The Scientific and Technological Research Council of Turkey (TUBITAK) in the framework of BIDEB-2219 International Postdoctoral Research Fellowship Program. K. Azizi is thankful to Iran Science Elites Federation (Saramadan)
for the partial  financial support provided under Grant No ISEF/M/400150.
\end{acknowledgments}

\bibliography{deltazerodibaryon}

\begin{thebibliography}{10}
\expandafter\ifx\csname url\endcsname\relax
  \def\url#1{\texttt{#1}}\fi
\expandafter\ifx\csname urlprefix\endcsname\relax\def\urlprefix{URL }\fi
\expandafter\ifx\csname href\endcsname\relax
  \def\href#1#2{#2} \def\path#1{#1}\fi

\bibitem{Gell-Mann:1964ewy}
M.~Gell-Mann, {A Schematic Model of Baryons and Mesons}, Phys. Lett. 8 (1964)
  214--215.
\newblock \href {http://dx.doi.org/10.1016/S0031-9163(64)92001-3}
  {\path{doi:10.1016/S0031-9163(64)92001-3}}.

\bibitem{Chen:2016qju}
H.-X. Chen, W.~Chen, X.~Liu, S.-L. Zhu, {The hidden-charm pentaquark and
  tetraquark states}, Phys. Rept. 639 (2016) 1--121.
\newblock \href {http://arxiv.org/abs/1601.02092} {\path{arXiv:1601.02092}},
  \href {http://dx.doi.org/10.1016/j.physrep.2016.05.004}
  {\path{doi:10.1016/j.physrep.2016.05.004}}.

\bibitem{Richard:2016eis}
J.-M. Richard, {Exotic hadrons: review and perspectives}, Few Body Syst.
  57~(12) (2016) 1185--1212.
\newblock \href {http://arxiv.org/abs/1606.08593} {\path{arXiv:1606.08593}},
  \href {http://dx.doi.org/10.1007/s00601-016-1159-0}
  {\path{doi:10.1007/s00601-016-1159-0}}.

\bibitem{Lebed:2016hpi}
R.~F. Lebed, R.~E. Mitchell, E.~S. Swanson, {Heavy-Quark QCD Exotica}, Prog.
  Part. Nucl. Phys. 93 (2017) 143--194.
\newblock \href {http://arxiv.org/abs/1610.04528} {\path{arXiv:1610.04528}},
  \href {http://dx.doi.org/10.1016/j.ppnp.2016.11.003}
  {\path{doi:10.1016/j.ppnp.2016.11.003}}.

\bibitem{Esposito:2016noz}
A.~Esposito, A.~Pilloni, A.~D. Polosa, {Multiquark Resonances}, Phys. Rept. 668
  (2017) 1--97.
\newblock \href {http://arxiv.org/abs/1611.07920} {\path{arXiv:1611.07920}},
  \href {http://dx.doi.org/10.1016/j.physrep.2016.11.002}
  {\path{doi:10.1016/j.physrep.2016.11.002}}.

\bibitem{Guo:2017jvc}
F.-K. Guo, C.~Hanhart, U.-G. Mei\ss{}ner, Q.~Wang, Q.~Zhao, B.-S. Zou,
  {Hadronic molecules}, Rev. Mod. Phys. 90~(1) (2018) 015004.
\newblock \href {http://arxiv.org/abs/1705.00141} {\path{arXiv:1705.00141}},
  \href {http://dx.doi.org/10.1103/RevModPhys.90.015004}
  {\path{doi:10.1103/RevModPhys.90.015004}}.

\bibitem{Ali:2017jda}
A.~Ali, J.~S. Lange, S.~Stone, {Exotics: Heavy Pentaquarks and Tetraquarks},
  Prog. Part. Nucl. Phys. 97 (2017) 123--198.
\newblock \href {http://arxiv.org/abs/1706.00610} {\path{arXiv:1706.00610}},
  \href {http://dx.doi.org/10.1016/j.ppnp.2017.08.003}
  {\path{doi:10.1016/j.ppnp.2017.08.003}}.

\bibitem{Olsen:2017bmm}
S.~L. Olsen, T.~Skwarnicki, D.~Zieminska, {Nonstandard heavy mesons and
  baryons: Experimental evidence}, Rev. Mod. Phys. 90~(1) (2018) 015003.
\newblock \href {http://arxiv.org/abs/1708.04012} {\path{arXiv:1708.04012}},
  \href {http://dx.doi.org/10.1103/RevModPhys.90.015003}
  {\path{doi:10.1103/RevModPhys.90.015003}}.

\bibitem{Liu:2019zoy}
Y.-R. Liu, H.-X. Chen, W.~Chen, X.~Liu, S.-L. Zhu, {Pentaquark and Tetraquark
  states}, Prog. Part. Nucl. Phys. 107 (2019) 237--320.
\newblock \href {http://arxiv.org/abs/1903.11976} {\path{arXiv:1903.11976}},
  \href {http://dx.doi.org/10.1016/j.ppnp.2019.04.003}
  {\path{doi:10.1016/j.ppnp.2019.04.003}}.

\bibitem{Brambilla:2019esw}
N.~Brambilla, S.~Eidelman, C.~Hanhart, A.~Nefediev, C.-P. Shen, C.~E. Thomas,
  A.~Vairo, C.-Z. Yuan, {The $XYZ$ states: experimental and theoretical status
  and perspectives}, Phys. Rept. 873 (2020) 1--154.
\newblock \href {http://arxiv.org/abs/1907.07583} {\path{arXiv:1907.07583}},
  \href {http://dx.doi.org/10.1016/j.physrep.2020.05.001}
  {\path{doi:10.1016/j.physrep.2020.05.001}}.

\bibitem{Agaev:2020zad}
S.~Agaev, K.~Azizi, H.~Sundu, {Four-quark exotic mesons}, Turk. J. Phys. 44~(2)
  (2020) 95--173.
\newblock \href {http://arxiv.org/abs/2004.12079} {\path{arXiv:2004.12079}},
  \href {http://dx.doi.org/10.3906/fiz-2003-15}
  {\path{doi:10.3906/fiz-2003-15}}.

\bibitem{WASA-at-COSY:2011bjg}
P.~Adlarson, et~al., {ABC Effect in Basic Double-Pionic Fusion --- Observation
  of a new resonance?}, Phys. Rev. Lett. 106 (2011) 242302.
\newblock \href {http://arxiv.org/abs/1104.0123} {\path{arXiv:1104.0123}},
  \href {http://dx.doi.org/10.1103/PhysRevLett.106.242302}
  {\path{doi:10.1103/PhysRevLett.106.242302}}.

\bibitem{WASA-at-COSY:2012seb}
P.~Adlarson, et~al., {Isospin Decomposition of the Basic Double-Pionic Fusion
  in the Region of the ABC Effect}, Phys. Lett. B 721 (2013) 229--236.
\newblock \href {http://arxiv.org/abs/1212.2881} {\path{arXiv:1212.2881}},
  \href {http://dx.doi.org/10.1016/j.physletb.2013.03.019}
  {\path{doi:10.1016/j.physletb.2013.03.019}}.

\bibitem{WASA-at-COSY:2013fzt}
P.~Adlarson, et~al., {Measurement of the $pn \to pp\pi^0\pi^-$ reaction in
  search for the recently observed resonance structure in $d\pi^0\pi^0$ and
  $d\pi^+\pi^-$ systems}, Phys. Rev. C 88~(5) (2013) 055208.
\newblock \href {http://arxiv.org/abs/1306.5130} {\path{arXiv:1306.5130}},
  \href {http://dx.doi.org/10.1103/PhysRevC.88.055208}
  {\path{doi:10.1103/PhysRevC.88.055208}}.

\bibitem{WASA-at-COSY:2014dmv}
P.~Adlarson, et~al., {Evidence for a New Resonance from Polarized
  Neutron-Proton Scattering}, Phys. Rev. Lett. 112~(20) (2014) 202301.
\newblock \href {http://arxiv.org/abs/1402.6844} {\path{arXiv:1402.6844}},
  \href {http://dx.doi.org/10.1103/PhysRevLett.112.202301}
  {\path{doi:10.1103/PhysRevLett.112.202301}}.

\bibitem{Dyson:1964xwa}
F.~Dyson, N.~H. Xuong, {Y=2 States in Su(6) Theory}, Phys. Rev. Lett. 13~(26)
  (1964) 815--817.
\newblock \href {http://dx.doi.org/10.1103/PhysRevLett.13.815}
  {\path{doi:10.1103/PhysRevLett.13.815}}.

\bibitem{Ikeno:2021frl}
N.~Ikeno, R.~Molina, E.~Oset, {Triangle singularity mechanism for the $
  pp\rightarrow \pi+d$ fusion reaction}, Phys. Rev. C 104~(1) (2021) 014614.
\newblock \href {http://arxiv.org/abs/2103.01712} {\path{arXiv:2103.01712}},
  \href {http://dx.doi.org/10.1103/PhysRevC.104.014614}
  {\path{doi:10.1103/PhysRevC.104.014614}}.

\bibitem{Molina:2021bwp}
R.~Molina, N.~Ikeno, E.~Oset, {Sequential single pion production explaning the
  dibaryon ''$d^*(2380)$'' peak}\href {http://arxiv.org/abs/2102.05575}
  {\path{arXiv:2102.05575}}.

\bibitem{Jaffe:1976yi}
R.~L. Jaffe, {Perhaps a Stable Dihyperon}, Phys. Rev. Lett. 38 (1977) 195--198,
  [Erratum: Phys.Rev.Lett. 38, 617 (1977)].
\newblock \href {http://dx.doi.org/10.1103/PhysRevLett.38.195}
  {\path{doi:10.1103/PhysRevLett.38.195}}.

\bibitem{Liu:1982wg}
K.~F. Liu, C.~W. Wong, {MIT BAG MODEL WITH CENTER-OF-MASS CORRECTION}, Phys.
  Lett. B 113 (1982) 1--5.
\newblock \href {http://dx.doi.org/10.1016/0370-2693(82)90096-X}
  {\path{doi:10.1016/0370-2693(82)90096-X}}.

\bibitem{Yost:1985mj}
S.~A. Yost, C.~R. Nappi, {The Mass of the $H$ Dibaryon in a Chiral Model},
  Phys. Rev. D 32 (1985) 816.
\newblock \href {http://dx.doi.org/10.1103/PhysRevD.32.816}
  {\path{doi:10.1103/PhysRevD.32.816}}.

\bibitem{Oka:1983ku}
M.~Oka, K.~Shimizu, K.~Yazaki, {The Dihyperon State in the Quark Cluster
  Model}, Phys. Lett. B 130 (1983) 365--368.
\newblock \href {http://dx.doi.org/10.1016/0370-2693(83)91523-X}
  {\path{doi:10.1016/0370-2693(83)91523-X}}.

\bibitem{Balachandran:1985fb}
A.~P. Balachandran, F.~Lizzi, V.~G.~J. Rodgers, A.~Stern, {Dibaryons as Chiral
  Solitons}, Nucl. Phys. B 256 (1985) 525--556.
\newblock \href {http://dx.doi.org/10.1016/0550-3213(85)90407-9}
  {\path{doi:10.1016/0550-3213(85)90407-9}}.

\bibitem{Oka:1986fr}
M.~Oka, K.~Shimizu, K.~Yazaki, {Hyperon - Nucleon and Hyperon-hyperon
  Interaction in a Quark Model}, Nucl. Phys. A 464 (1987) 700--716.
\newblock \href {http://dx.doi.org/10.1016/0375-9474(87)90371-X}
  {\path{doi:10.1016/0375-9474(87)90371-X}}.

\bibitem{Straub:1988mz}
U.~Straub, Z.-Y. Zhang, K.~Brauer, A.~Faessler, S.~B. Khadkikar, {Binding
  Energy of the Dihyperon in the Quark Cluster Model}, Phys. Lett. B 200 (1988)
  241--245.
\newblock \href {http://dx.doi.org/10.1016/0370-2693(88)90763-0}
  {\path{doi:10.1016/0370-2693(88)90763-0}}.

\bibitem{Koike:1989ak}
Y.~Koike, K.~Shimizu, K.~Yazaki, {Study of Hyperon - Nucleon and
  Hyperon-hyperon Interaction in the Flipflop Model}, Nucl. Phys. A 513 (1990)
  653--666.
\newblock \href {http://dx.doi.org/10.1016/0375-9474(90)90403-9}
  {\path{doi:10.1016/0375-9474(90)90403-9}}.

\bibitem{Larin:1985yt}
S.~A. Larin, V.~A. Matveev, A.~A. Ovchinnikov, A.~A. Pivovarov, {Determination
  of the mass of the Lambda Lambda dibaryon by the method of QCD sum rules},
  Sov. J. Nucl. Phys. 44 (1986) 690.
\newblock \href {http://arxiv.org/abs/hep-ph/0405035}
  {\path{arXiv:hep-ph/0405035}}.

\bibitem{Kodama:1994np}
N.~Kodama, M.~Oka, T.~Hatsuda, {H dibaryon in the QCD sum rule}, Nucl. Phys. A
  580 (1994) 445--454.
\newblock \href {http://arxiv.org/abs/hep-ph/9404221}
  {\path{arXiv:hep-ph/9404221}}, \href
  {http://dx.doi.org/10.1016/0375-9474(94)90908-3}
  {\path{doi:10.1016/0375-9474(94)90908-3}}.

\bibitem{Inoue:2010es}
T.~Inoue, N.~Ishii, S.~Aoki, T.~Doi, T.~Hatsuda, Y.~Ikeda, K.~Murano,
  H.~Nemura, K.~Sasaki, {Bound H-dibaryon in Flavor SU(3) Limit of Lattice
  QCD}, Phys. Rev. Lett. 106 (2011) 162002.
\newblock \href {http://arxiv.org/abs/1012.5928} {\path{arXiv:1012.5928}},
  \href {http://dx.doi.org/10.1103/PhysRevLett.106.162002}
  {\path{doi:10.1103/PhysRevLett.106.162002}}.

\bibitem{Leandri:1997ge}
J.~Leandri, B.~Silvestre-Brac, {Study of dibaryon states containing three
  different types of quarks}, Few Body Syst. 23 (1998) 39--51.
\newblock \href {http://dx.doi.org/10.1007/s006010050063}
  {\path{doi:10.1007/s006010050063}}.

\bibitem{Vijande:2011im}
J.~Vijande, A.~Valcarce, J.~M. Richard, {Stability of hexaquarks in the string
  limit of confinement}, Phys. Rev. D 85 (2012) 014019.
\newblock \href {http://arxiv.org/abs/1111.5921} {\path{arXiv:1111.5921}},
  \href {http://dx.doi.org/10.1103/PhysRevD.85.014019}
  {\path{doi:10.1103/PhysRevD.85.014019}}.

\bibitem{Azizi:2019xla}
K.~Azizi, S.~S. Agaev, H.~Sundu, {The Scalar Hexaquark $uuddss$: a Candidate to
  Dark Matter?}, J. Phys. G 47~(9) (2020) 095001.
\newblock \href {http://arxiv.org/abs/1904.09913} {\path{arXiv:1904.09913}},
  \href {http://dx.doi.org/10.1088/1361-6471/ab9a0e}
  {\path{doi:10.1088/1361-6471/ab9a0e}}.

\bibitem{Beiming:2021bkj}
C.~Beiming, J.~Gr\"onroos, T.~Ohlsson, {Phenomenological mass model for exotic
  hadrons and predictions for masses of non-strange dibaryons as hexaquarks},
  Nucl. Phys. B 974 (2022) 115616.
\newblock \href {http://arxiv.org/abs/2106.11978} {\path{arXiv:2106.11978}},
  \href {http://dx.doi.org/10.1016/j.nuclphysb.2021.115616}
  {\path{doi:10.1016/j.nuclphysb.2021.115616}}.

\bibitem{Gongyo:2017fjb}
S.~Gongyo, et~al., {Most Strange Dibaryon from Lattice QCD}, Phys. Rev. Lett.
  120~(21) (2018) 212001.
\newblock \href {http://arxiv.org/abs/1709.00654} {\path{arXiv:1709.00654}},
  \href {http://dx.doi.org/10.1103/PhysRevLett.120.212001}
  {\path{doi:10.1103/PhysRevLett.120.212001}}.

\bibitem{Morita:2019rph}
K.~Morita, S.~Gongyo, T.~Hatsuda, T.~Hyodo, Y.~Kamiya, A.~Ohnishi, {Probing
  $\Omega\Omega$ and $p\Omega$ dibaryons with femtoscopic correlations in
  relativistic heavy-ion collisions}, Phys. Rev. C 101~(1) (2020) 015201.
\newblock \href {http://arxiv.org/abs/1908.05414} {\path{arXiv:1908.05414}},
  \href {http://dx.doi.org/10.1103/PhysRevC.101.015201}
  {\path{doi:10.1103/PhysRevC.101.015201}}.

\bibitem{Huang:2019hmq}
H.~Huang, X.~Zhu, J.~Ping, {Reanalysis of the most strange dibaryon within
  constituent quark models}, Phys. Rev. C 101~(3) (2020) 034004.
\newblock \href {http://arxiv.org/abs/1912.11256} {\path{arXiv:1912.11256}},
  \href {http://dx.doi.org/10.1103/PhysRevC.101.034004}
  {\path{doi:10.1103/PhysRevC.101.034004}}.

\bibitem{Chen:2019vdh}
X.-H. Chen, Q.-N. Wang, W.~Chen, H.-X. Chen, {Exotic $\Omega\Omega$ dibaryon
  states in a molecular picture}, Chin. Phys. C 45~(4) (2021) 041002.
\newblock \href {http://arxiv.org/abs/1906.11089} {\path{arXiv:1906.11089}},
  \href {http://dx.doi.org/10.1088/1674-1137/abdfbe}
  {\path{doi:10.1088/1674-1137/abdfbe}}.

\bibitem{Haidenbauer:2021smk}
J.~Haidenbauer, U.-G. Mei\ss{}ner, {On the structure in the
  \ensuremath{\Lambda}N cross section at the \ensuremath{\Sigma}N threshold},
  Chin. Phys. C 45~(9) (2021) 094104.
\newblock \href {http://arxiv.org/abs/2105.00836} {\path{arXiv:2105.00836}},
  \href {http://dx.doi.org/10.1088/1674-1137/ac0e89}
  {\path{doi:10.1088/1674-1137/ac0e89}}.

\bibitem{HALQCD:2018qyu}
T.~Iritani, et~al., {$N\Omega$ dibaryon from lattice QCD near the physical
  point}, Phys. Lett. B 792 (2019) 284--289.
\newblock \href {http://arxiv.org/abs/1810.03416} {\path{arXiv:1810.03416}},
  \href {http://dx.doi.org/10.1016/j.physletb.2019.03.050}
  {\path{doi:10.1016/j.physletb.2019.03.050}}.

\bibitem{Huang:2019esu}
H.~Huang, J.~Ping, F.~Wang, {Prediction of $N\Omega$-like dibaryons with heavy
  quarks}, Phys. Rev. C 101~(1) (2020) 015204.
\newblock \href {http://arxiv.org/abs/1910.14277} {\path{arXiv:1910.14277}},
  \href {http://dx.doi.org/10.1103/PhysRevC.101.015204}
  {\path{doi:10.1103/PhysRevC.101.015204}}.

\bibitem{Chen:2021hxs}
X.-H. Chen, Q.-N. Wang, W.~Chen, H.-X. Chen, {Mass spectra of $N\Omega$
  dibaryons in the $^3S_1$ and $^5S_2$ channels}, Phys. Rev. D 103~(9) (2021)
  094011.
\newblock \href {http://arxiv.org/abs/2103.09739} {\path{arXiv:2103.09739}},
  \href {http://dx.doi.org/10.1103/PhysRevD.103.094011}
  {\path{doi:10.1103/PhysRevD.103.094011}}.

\bibitem{HALQCD:2019wsz}
K.~Sasaki, et~al., {$\Lambda\Lambda$ and N$\Xi$ interactions from lattice QCD
  near the physical point}, Nucl. Phys. A 998 (2020) 121737.
\newblock \href {http://arxiv.org/abs/1912.08630} {\path{arXiv:1912.08630}},
  \href {http://dx.doi.org/10.1016/j.nuclphysa.2020.121737}
  {\path{doi:10.1016/j.nuclphysa.2020.121737}}.

\bibitem{Wang:2017sto}
Z.-G. Wang, {Analysis of the scalar doubly charmed hexaquark state with QCD sum
  rules}, Eur. Phys. J. C 77~(9) (2017) 642.
\newblock \href {http://arxiv.org/abs/1707.09767} {\path{arXiv:1707.09767}},
  \href {http://dx.doi.org/10.1140/epjc/s10052-017-5207-9}
  {\path{doi:10.1140/epjc/s10052-017-5207-9}}.

\bibitem{Wan:2019ake}
B.-D. Wan, L.~Tang, C.-F. Qiao, {Hidden-bottom and -charm hexaquark states in
  QCD sum rules}, Eur. Phys. J. C 80~(2) (2020) 121.
\newblock \href {http://arxiv.org/abs/1912.12046} {\path{arXiv:1912.12046}},
  \href {http://dx.doi.org/10.1140/epjc/s10052-020-7701-8}
  {\path{doi:10.1140/epjc/s10052-020-7701-8}}.

\bibitem{Wang:2019gal}
Z.-G. Wang, {Triply-charmed dibaryon states or two-baryon scattering states
  from QCD sum rules}, Phys. Rev. D 102~(3) (2020) 034008.
\newblock \href {http://arxiv.org/abs/1912.07230} {\path{arXiv:1912.07230}},
  \href {http://dx.doi.org/10.1103/PhysRevD.102.034008}
  {\path{doi:10.1103/PhysRevD.102.034008}}.

\bibitem{Junnarkar:2019equ}
P.~Junnarkar, N.~Mathur, {Deuteronlike Heavy Dibaryons from Lattice Quantum
  Chromodynamics}, Phys. Rev. Lett. 123~(16) (2019) 162003.
\newblock \href {http://arxiv.org/abs/1906.06054} {\path{arXiv:1906.06054}},
  \href {http://dx.doi.org/10.1103/PhysRevLett.123.162003}
  {\path{doi:10.1103/PhysRevLett.123.162003}}.

\bibitem{Pan:2020xek}
Y.-W. Pan, M.-Z. Liu, L.-S. Geng, {Triply charmed dibaryons in the one boson
  exchange model}, Phys. Rev. D 102~(5) (2020) 054025.
\newblock \href {http://arxiv.org/abs/2004.07467} {\path{arXiv:2004.07467}},
  \href {http://dx.doi.org/10.1103/PhysRevD.102.054025}
  {\path{doi:10.1103/PhysRevD.102.054025}}.

\bibitem{Wang:2020jqu}
Z.-G. Wang, {Triply-charmed hexaquark states with the QCD sum rules}, Int. J.
  Mod. Phys. A 35~(14) (2020) 2050073.
\newblock \href {http://arxiv.org/abs/2002.06202} {\path{arXiv:2002.06202}},
  \href {http://dx.doi.org/10.1142/S0217751X20500736}
  {\path{doi:10.1142/S0217751X20500736}}.

\bibitem{Richard:2020zxb}
J.-M. Richard, A.~Valcarce, J.~Vijande, {Very heavy flavored dibaryons}, Phys.
  Rev. Lett. 124~(21) (2020) 212001.
\newblock \href {http://arxiv.org/abs/2005.06894} {\path{arXiv:2005.06894}},
  \href {http://dx.doi.org/10.1103/PhysRevLett.124.212001}
  {\path{doi:10.1103/PhysRevLett.124.212001}}.

\bibitem{Huang:2020bmb}
H.~Huang, J.~Ping, X.~Zhu, F.~Wang, {Full heavy dibaryons}\href
  {http://arxiv.org/abs/2011.00513} {\path{arXiv:2011.00513}}.

\bibitem{Lyu:2021qsh}
Y.~Lyu, H.~Tong, T.~Sugiura, S.~Aoki, T.~Doi, T.~Hatsuda, J.~Meng, T.~Miyamoto,
  {Dibaryon with Highest Charm Number near Unitarity from Lattice QCD}, Phys.
  Rev. Lett. 127~(7) (2021) 072003.
\newblock \href {http://arxiv.org/abs/2102.00181} {\path{arXiv:2102.00181}},
  \href {http://dx.doi.org/10.1103/PhysRevLett.127.072003}
  {\path{doi:10.1103/PhysRevLett.127.072003}}.

\bibitem{Xia:2021nif}
Z.~Xia, S.~Fan, X.~Zhu, H.~Huang, J.~Ping, {Search for doubly heavy dibaryons
  in the quark delocalization color screening model}, Phys. Rev. C 105~(2)
  (2022) 025201.
\newblock \href {http://arxiv.org/abs/2105.14723} {\path{arXiv:2105.14723}},
  \href {http://dx.doi.org/10.1103/PhysRevC.105.025201}
  {\path{doi:10.1103/PhysRevC.105.025201}}.

\bibitem{Liu:2021pdu}
M.-Z. Liu, L.-S. Geng, {Prediction of an
  \ensuremath{\Omega}bbb\ensuremath{\Omega}bbb Dibaryon in the Extended
  One-Boson Exchange Model}, Chin. Phys. Lett. 38~(10) (2021) 101201.
\newblock \href {http://arxiv.org/abs/2107.04957} {\path{arXiv:2107.04957}},
  \href {http://dx.doi.org/10.1088/0256-307X/38/10/101201}
  {\path{doi:10.1088/0256-307X/38/10/101201}}.

\bibitem{Fritzsch:1972jv}
H.~Fritzsch, M.~Gell-Mann, {Current algebra: Quarks and what else?}, eConf
  C720906V2 (1972) 135--165.
\newblock \href {http://arxiv.org/abs/hep-ph/0208010}
  {\path{arXiv:hep-ph/0208010}}.

\bibitem{Dai:2005kt}
L.-R. Dai, {Delta Delta dibaryon structure in extended chiral SU(3) quark
  model}, Chin. Phys. Lett. 22 (2005) 2204--2206.
\newblock \href {http://arxiv.org/abs/nucl-th/0507042}
  {\path{arXiv:nucl-th/0507042}}, \href
  {http://dx.doi.org/10.1088/0256-307X/22/9/018}
  {\path{doi:10.1088/0256-307X/22/9/018}}.

\bibitem{Huang:2013nba}
H.~Huang, J.~Ping, F.~Wang, {Dynamical calculation of the $\Delta\Delta$
  dibaryon candidates}, Phys. Rev. C 89~(3) (2014) 034001.
\newblock \href {http://arxiv.org/abs/1312.7756} {\path{arXiv:1312.7756}},
  \href {http://dx.doi.org/10.1103/PhysRevC.89.034001}
  {\path{doi:10.1103/PhysRevC.89.034001}}.

\bibitem{Gal:2013dca}
A.~Gal, H.~Garcilazo, {Three-Body Calculation of the Delta-Delta Dibaryon
  Candidate D(03) at 2.37 GeV}, Phys. Rev. Lett. 111 (2013) 172301.
\newblock \href {http://arxiv.org/abs/1308.2112} {\path{arXiv:1308.2112}},
  \href {http://dx.doi.org/10.1103/PhysRevLett.111.172301}
  {\path{doi:10.1103/PhysRevLett.111.172301}}.

\bibitem{Shifman:1978bx}
M.~A. Shifman, A.~I. Vainshtein, V.~I. Zakharov, {QCD and Resonance Physics.
  Theoretical Foundations}, Nucl. Phys. B 147 (1979) 385--447.
\newblock \href {http://dx.doi.org/10.1016/0550-3213(79)90022-1}
  {\path{doi:10.1016/0550-3213(79)90022-1}}.

\bibitem{Shifman:1978by}
M.~A. Shifman, A.~I. Vainshtein, V.~I. Zakharov, {QCD and Resonance Physics:
  Applications}, Nucl. Phys. B 147 (1979) 448--518.
\newblock \href {http://dx.doi.org/10.1016/0550-3213(79)90023-3}
  {\path{doi:10.1016/0550-3213(79)90023-3}}.

\bibitem{Ioffe:1981kw}
B.~L. Ioffe, {Calculation of Baryon Masses in Quantum Chromodynamics}, Nucl.
  Phys. B 188 (1981) 317--341, [Erratum: Nucl.Phys.B 191, 591--592 (1981)].
\newblock \href {http://dx.doi.org/10.1016/0550-3213(81)90259-5}
  {\path{doi:10.1016/0550-3213(81)90259-5}}.

\bibitem{Narison:1989aq}
S.~Narison, {QCD spectral sum rules}, Vol.~26, 1989.

\bibitem{Launer:1983ib}
G.~Launer, S.~Narison, R.~Tarrach, {Nonperturbative {QCD} Vacuum From $e^+ e^-
  \to$ I = 1 Hadron Data}, Z. Phys. C 26 (1984) 433--439.
\newblock \href {http://dx.doi.org/10.1007/BF01452571}
  {\path{doi:10.1007/BF01452571}}.

\bibitem{Narison:2009vy}
S.~Narison, {Power corrections to alpha(s)(M(tau)),|V(us)| and anti-m(s)},
  Phys. Lett. B 673 (2009) 30--36.
\newblock \href {http://arxiv.org/abs/0901.3823} {\path{arXiv:0901.3823}},
  \href {http://dx.doi.org/10.1016/j.physletb.2009.01.062}
  {\path{doi:10.1016/j.physletb.2009.01.062}}.

\bibitem{Braghin:2014nva}
F.~L. Braghin, F.~S. Navarra, {Factorization breaking of four-quark condensates
  in the Nambu\textendash{}Jona-Lasinio model}, Phys. Rev. D 91~(7) (2015)
  074008.
\newblock \href {http://arxiv.org/abs/1404.4094} {\path{arXiv:1404.4094}},
  \href {http://dx.doi.org/10.1103/PhysRevD.91.074008}
  {\path{doi:10.1103/PhysRevD.91.074008}}.

\bibitem{Albuquerque:2010fm}
R.~M. Albuquerque, J.~M. Dias, M.~Nielsen, {Can the X(4350) narrow structure be
  a $1^{-+}$ exotic state?}, Phys. Lett. B 690 (2010) 141--144.
\newblock \href {http://arxiv.org/abs/1001.3092} {\path{arXiv:1001.3092}},
  \href {http://dx.doi.org/10.1016/j.physletb.2010.05.024}
  {\path{doi:10.1016/j.physletb.2010.05.024}}.

\bibitem{Albuquerque:2021erv}
R.~M. Albuquerque, S.~Narison, D.~Rabetiarivony, G.~Randriamanatrika, {Doubly
  hidden 0++ molecules and tetraquarks states from QCD at NLO}, Nucl. Part.
  Phys. Proc. 312-317 (2021) 120--124.
\newblock \href {http://arxiv.org/abs/2102.08776} {\path{arXiv:2102.08776}},
  \href {http://dx.doi.org/10.1016/j.nuclphysbps.2021.05.031}
  {\path{doi:10.1016/j.nuclphysbps.2021.05.031}}.

\bibitem{Albuquerque:2016znh}
R.~Albuquerque, S.~Narison, F.~Fanomezana, A.~Rabemananjara, D.~Rabetiarivony,
  G.~Randriamanatrika, {XYZ-like Spectra from Laplace Sum Rule at N2LO in the
  Chiral Limit}, Int. J. Mod. Phys. A 31~(36) (2016) 1650196.
\newblock \href {http://arxiv.org/abs/1609.03351} {\path{arXiv:1609.03351}},
  \href {http://dx.doi.org/10.1142/S0217751X16501967}
  {\path{doi:10.1142/S0217751X16501967}}.

\bibitem{Albuquerque:2017vfq}
R.~Albuquerque, S.~Narison, D.~Rabetiarivony, G.~Randriamanatrika, {XYZ-SU3
  Breakings from Laplace Sum Rules at Higher Orders}, Int. J. Mod. Phys. A
  33~(16) (2018) 1850082.
\newblock \href {http://arxiv.org/abs/1709.09023} {\path{arXiv:1709.09023}},
  \href {http://dx.doi.org/10.1142/S0217751X18500823}
  {\path{doi:10.1142/S0217751X18500823}}.

\bibitem{Belyaev:1982sa}
V.~M. Belyaev, B.~L. Ioffe, {Determination of Baryon and Baryonic Resonance
  Masses from QCD Sum Rules. 1. Nonstrange Baryons}, Sov. Phys. JETP 56 (1982)
  493--501.

\bibitem{Belyaev:1982cd}
V.~M. Belyaev, B.~L. Ioffe, {Determination of the baryon mass and baryon
  resonances from the quantum-chromodynamics sum rule. Strange baryons}, Sov.
  Phys. JETP 57 (1983) 716--721.

\bibitem{Chen:2014vha}
H.-X. Chen, E.-L. Cui, W.~Chen, T.~G. Steele, S.-L. Zhu, {QCD sum rule study of
  the d*(2380)}, Phys. Rev. C 91~(2) (2015) 025204.
\newblock \href {http://arxiv.org/abs/1410.0394} {\path{arXiv:1410.0394}},
  \href {http://dx.doi.org/10.1103/PhysRevC.91.025204}
  {\path{doi:10.1103/PhysRevC.91.025204}}.

\bibitem{Krasnikov:1982ea}
N.~V. Krasnikov, A.~A. Pivovarov, N.~N. Tavkhelidze, {The Use of Finite Energy
  Sum Rules for the Description of the Hadronic Properties of QCD}, Z. Phys. C
  19 (1983) 301.
\newblock \href {http://dx.doi.org/10.1007/BF01577186}
  {\path{doi:10.1007/BF01577186}}.

\bibitem{Jamin:1987gq}
M.~Jamin, {Radiative Corrections for Baryonic Correlators}, Z. Phys. C 37
  (1988) 635.
\newblock \href {http://dx.doi.org/10.1007/BF01549725}
  {\path{doi:10.1007/BF01549725}}.

\bibitem{Groote:1999zp}
S.~Groote, J.~G. Korner, A.~A. Pivovarov, {O(alpha(s)) corrections to the
  correlator of finite mass baryon currents}, Phys. Rev. D 61 (2000) 071501.
\newblock \href {http://arxiv.org/abs/hep-ph/9911393}
  {\path{arXiv:hep-ph/9911393}}, \href
  {http://dx.doi.org/10.1103/PhysRevD.61.071501}
  {\path{doi:10.1103/PhysRevD.61.071501}}.

\bibitem{Groote:2014pva}
S.~Groote, J.~G. K\"orner, D.~Niinepuu, {Perturbative $O(\alpha_s)$ corrections
  to the correlation functions of light tetraquark currents}, Phys. Rev. D
  90~(5) (2014) 054028.
\newblock \href {http://arxiv.org/abs/1401.4801} {\path{arXiv:1401.4801}},
  \href {http://dx.doi.org/10.1103/PhysRevD.90.054028}
  {\path{doi:10.1103/PhysRevD.90.054028}}.

\bibitem{Groote:2006sy}
S.~Groote, J.~G. Korner, A.~A. Pivovarov, {Large next-to-leading order QCD
  corrections to pentaquark sum rules}, Phys. Rev. D 74 (2006) 017503.
\newblock \href {http://arxiv.org/abs/hep-ph/0603220}
  {\path{arXiv:hep-ph/0603220}}, \href
  {http://dx.doi.org/10.1103/PhysRevD.74.017503}
  {\path{doi:10.1103/PhysRevD.74.017503}}.

\bibitem{Groote:2011my}
S.~Groote, J.~G. Korner, A.~A. Pivovarov, {Calculating Loops without Loop
  Calculations: NLO Computation of Pentaquark Correlators}, Phys. Rev. D 86
  (2012) 034023.
\newblock \href {http://arxiv.org/abs/1107.0615} {\path{arXiv:1107.0615}},
  \href {http://dx.doi.org/10.1103/PhysRevD.86.034023}
  {\path{doi:10.1103/PhysRevD.86.034023}}.

\bibitem{Wang:2021qmn}
X.-W. Wang, Z.-G. Wang, G.-l. Yu, {Study of $\Lambda _c\Lambda _c$ dibaryon and
  $\Lambda _c{\bar{\Lambda }}_c$ baryonium states via QCD sum rules}, Eur.
  Phys. J. A 57~(9) (2021) 275.
\newblock \href {http://arxiv.org/abs/2107.04751} {\path{arXiv:2107.04751}},
  \href {http://dx.doi.org/10.1140/epja/s10050-021-00576-8}
  {\path{doi:10.1140/epja/s10050-021-00576-8}}.

\bibitem{Zyla:2020zbs}
P.~A. Zyla, et~al., {Review of Particle Physics}, PTEP 2020~(8) (2020) 083C01.
\newblock \href {http://dx.doi.org/10.1093/ptep/ptaa104}
  {\path{doi:10.1093/ptep/ptaa104}}.

\end{thebibliography}

\end{document}